\documentclass[12pt]{article} 
\usepackage[top=30mm, bottom=40mm, left=25mm, right=25mm]{geometry}
\usepackage{cite} 

\usepackage{graphicx}
\usepackage{amsmath, amssymb} 				
\usepackage{empheq}
\usepackage{fancybox}
\usepackage{mathrsfs} 
\usepackage{textcomp} 
\usepackage{mathtools}

\usepackage{epstopdf}

\usepackage{latexsym}
\usepackage{marvosym}
\usepackage{bm}	

\usepackage[mathcal]{euscript}
\usepackage{slashed}
\usepackage[all]{xy}
\usepackage{tikz}
	\usetikzlibrary{decorations.markings}
	\usetikzlibrary{arrows}
\usepackage{indentfirst}

\numberwithin{equation}{section}

\usepackage{hyperref}
\hypersetup{
	unicode,	
	colorlinks,
	citecolor=cyan,
	linkcolor=cyan, 
	bookmarksopen=true,
	bookmarksopenlevel=\maxdimen,
	}

\usepackage{amsthm} 
\theoremstyle{definition}

\theoremstyle{plain}

\def\K3{\mathrm K3}

\def\U#1{U({#1})}

\def\double #1{#1{\hbox{\kern-2pt $#1$}}}

\def\dd{\hbox{\,\Large$\triangleright$}}

\def\pp{{\mathchoice
          {
              \kern 1pt%
              \raise 1pt
              \vbox{\hrule width5pt height0.4pt depth0pt
                    \kern -2pt
                    \hbox{\kern 2.3pt
                          \vrule width0.4pt height6pt depth0pt
                          }
                    \kern -2pt
                    \hrule width5pt height0.4pt depth0pt}%
                    \kern 1pt
           }
            {
              \kern 1pt%
              \raise 1pt
              \vbox{\hrule width4.3pt height0.4pt depth0pt
                    \kern -1.8pt
                    \hbox{\kern 1.95pt
                          \vrule width0.4pt height5.4pt depth0pt
                          }
                    \kern -1.8pt
                    \hrule width4.3pt height0.4pt depth0pt}%
                    \kern 1pt
            }
            {
              \kern 0.5pt%
              \raise 1pt
              \vbox{\hrule width4.0pt height0.3pt depth0pt
                    \kern -1.9pt  
                    \hbox{\kern 1.85pt
                          \vrule width0.3pt height5.7pt depth0pt
                          }
                    \kern -1.9pt
                    \hrule width4.0pt height0.3pt depth0pt}%
                    \kern 0.5pt
            }
            {
              \kern 0.5pt%
              \raise 1pt
              \vbox{\hrule width3.6pt height0.3pt depth0pt
                    \kern -1.5pt
                    \hbox{\kern 1.65pt
                          \vrule width0.3pt height4.5pt depth0pt
                          }
                    \kern -1.5pt
                    \hrule width3.6pt height0.3pt depth0pt}%
                    \kern 0.5pt
            }
        }}

\def\mm{{\mathchoice
                       {
                             \kern 1pt
               \raise 1pt    \vbox{\hrule width5pt height0.4pt depth0pt
                                  \kern 2pt
                                  \hrule width5pt height0.4pt depth0pt}
                             \kern 1pt}
                       {
                            \kern 1pt
               \raise 1pt \vbox{\hrule width4.3pt height0.4pt depth0pt
                                  \kern 1.8pt
                                  \hrule width4.3pt height0.4pt depth0pt}
                             \kern 1pt}
                       {
                            \kern 0.5pt
               \raise 1pt
                            \vbox{\hrule width4.0pt height0.3pt depth0pt
                                  \kern 1.9pt
                                  \hrule width4.0pt height0.3pt depth0pt}
                            \kern 1pt}
                       {
                           \kern 0.5pt
             \raise 1pt  \vbox{\hrule width3.6pt height0.3pt depth0pt
                                  \kern 1.5pt
                                  \hrule width3.6pt height0.3pt depth0pt}
                           \kern 0.5pt}
                       }}

\def\ad{{\kern0.5pt
                   \alpha \kern-5.05pt
\raise5.8pt\hbox{$\textstyle.$}\kern 0.5pt}}

\def\bd{{\kern0.5pt
                   \beta \kern-5.05pt \raise5.8pt\hbox{$\textstyle.$}\kern 0.5pt}}

\def\qd{{\kern0.5pt
                   q \kern-5.05pt \raise5.8pt\hbox{$\textstyle.$}\kern 0.5pt}}
\def\Dot#1{{\kern0.5pt
     {#1} \kern-5.05pt \raise5.8pt\hbox{$\textstyle.$}\kern 0.5pt}}

\def\on#1#2{{\buildrel{\mkern2.5mu#1\mkern-2.5mu}\over{#2}}}
\def\dt#1{\on{\hbox{\bf .}}{#1}}                
\def\under#1#2{\mathop{\null#2}\limits_{#1}}	
\mathcode`\*="702A                  
\def\f#1#2{{\textstyle{#1\over#2}}}	   
\def\half{{\textstyle{1\over{\raise.1ex\hbox{$\scriptstyle{2}$}}}}}

\catcode128=13 \def €{\"A}                 
\catcode129=13 \def {\AA}                 
\catcode130=13 \def '{\c}           	   
\catcode131=13 \def ƒ{\'E}                   
\catcode132=13 \def "{\~N}                   
\catcode133=13 \def …{\"O}                 
\catcode134=13 \def †{\"U}                  
\catcode135=13 \def ‡{\'a}                  
\catcode136=13 \def ˆ{\`a}                   
\catcode137=13 \def ‰{\^a}                 
\catcode138=13 \def Š{\"a}                 
\catcode139=13 \def ‹{\~a}                   
\catcode140=13 \def Œ{\alpha}            
\catcode141=13 \def {\chi}                
\catcode142=13 \def Ž{\'e}                   
\catcode143=13 \def {\`e}                    
\catcode144=13 \def {\^e}                  
\catcode145=13 \def '{\"e}                
\catcode146=13 \def '{\'\i}                 
\catcode147=13 \def "{\`\i}                  
\catcode148=13 \def "{\^\i}                
\catcode149=13 \def •{\"\i}                
\catcode150=13 \def –{\~n}                  
\catcode151=13 \def —{\'o}                 
\catcode152=13 \def ˜{\`o}                  
\catcode153=13 \def ™{\^o}                
\catcode154=13 \def š{\"o}                 
\catcode155=13 \def ›{\~o}                  
\catcode156=13 \def œ{\'u}                  
\catcode157=13 \def {\`u}                  
\catcode158=13 \def ž{\^u}                
\catcode159=13 \def Ÿ{\"u}                
\catcode160=13 \def  {\tau}               
\catcode162=13 \def ¢{\oplus}           
\catcode163=13 \def £{\relax\ifmmode\expandafter\to\else\itemize\fi} 
\catcode164=13 \def ¤{\subset}	  
\catcode165=13 \def ¥{\infty}           
\catcode166=13 \def ¦{\mp}                
\catcode167=13 \def §{\sigma}           
\catcode168=13 \def ¨{\rho}               
\catcode169=13 \def ©{\gamma}         
\catcode170=13 \def ª{\leftrightarrow} 
\catcode171=13 \def «{\relax\ifmmode\expandafter\acute\else\expandafter\'\fi}
\catcode172=13 \def ¬{\relax\ifmmode\expandafter\ddt\else\expandafter\"\fi}
\catcode173=13 \def ­{\equiv}            
\catcode174=13 \def ®{{{}={}}}          
\catcode175=13 \def ¯{\Omega}          
\catcode176=13 \def °{\otimes}          
\catcode177=13 \def ±{\ne}                 
\catcode178=13 \def ²{\le}                   
\catcode179=13 \def ³{\ge}                  
\catcode180=13 \def ´{\upsilon}          
\catcode181=13 \def µ{\mu}                
\catcode182=13 \def ¶{\delta}             
\catcode183=13 \def ·{\epsilon}          
\catcode184=13 \def ¸{\Pi}                  
\catcode185=13 \def ¹{\pi}                  
\catcode186=13 \def º{\beta}               
\catcode187=13 \def »{\partial}           
\catcode188=13 \def ¼{\nobreak\ }       
\catcode189=13 \def ½{\zeta}               
\catcode190=13 \def ¾{\sim}                 
\catcode191=13 \def ¿{\omega}           
\catcode192=13 \def À{\dt}                     
\catcode193=13 \def Á{\gets}                
\catcode194=13 \def Â{\lambda}           
\catcode195=13 \def Ã{\nu}                   
\catcode196=13 \def Ä{\phi}                  
\catcode197=13 \def Å{\xi}                     
\catcode198=13 \def Æ{\psi}                  
\catcode199=13 \def Ç{\int}                    
\catcode200=13 \def È{\oint}                 
\catcode201=13 \def É{\relax\ifmmode\expandafter\cdot\else\vol\fi}    
\catcode202=13 \def Ê{\relax\ifmmode\expandafter\,\else\thinspace\fi}
\catcode203=13 \def Ë{\`A}                      
\catcode204=13 \def Ì{\~A}                      
\catcode205=13 \def Í{\~O}                      
\catcode206=13 \def Î{\Theta}              
\catcode207=13 \def Ï{\theta}               
\catcode208=13 \def Ð{\relax\ifmmode\expandafter\bar\else\expandafter\=\fi}
\catcode209=13 \def Ñ{\overline}             
\catcode210=13 \def Ò{\langle}               
\catcode211=13 \def Ó{\relax\ifmmode\expandafter\{\else\ital\fi}      
\catcode212=13 \def Ô{\rangle}               
\catcode213=13 \def Õ{\}}                        
\catcode214=13 \def Ö{\sla}                      
\catcode215=13 \def ×{\relax\ifmmode\expandafter\check\else\expandafter\v\fi}
\catcode216=13 \def Ø{\"y}                     
\catcode217=13 \def Ù{\"Y}  		    
\catcode218=13 \def Ú{\Leftarrow}       
\catcode219=13 \def Û{\Leftrightarrow}       
\catcode220=13 \def Ü{\relax\ifmmode\expandafter\Rightarrow\else\sect\fi}
\catcode221=13 \def Ý{\sum}                  
\catcode222=13 \def Þ{\prod}                 
\catcode223=13 \def ß{\widehat}              
\catcode224=13 \def à{\pm}                     
\catcode225=13 \def á{\nabla}                
\catcode226=13 \def â{\quad}                 
\catcode227=13 \def ã{\in}               	
\catcode228=13 \def ä{\star}      	      
\catcode229=13 \def å{\sqrt}                   
\catcode230=13 \def æ{\^E}			
\catcode231=13 \def ç{\Upsilon}              
\catcode232=13 \def è{\"E}    	   	 
\catcode233=13 \def é{\`E}               	  
\catcode234=13 \def ê{\Sigma}                
\catcode235=13 \def ë{\Delta}                 
\catcode236=13 \def ì{\Phi}                     
\catcode237=13 \def í{\`I}        		   
\catcode238=13 \def î{\iota}        	     
\catcode239=13 \def ï{\Psi}                     
\catcode240=13 \def ð{\times}                  
\catcode241=13 \def ñ{\Lambda}             
\catcode242=13 \def ò{\cdots}                
\catcode243=13 \def ó{\^U}			
\catcode244=13 \def ô{\`U}    	              
\catcode245=13 \def õ{\bo}                       
\catcode246=13 \def ö{\relax\ifmmode\expandafter\hat\else\expandafter\^\fi}
\catcode247=13 \def÷{\relax\ifmmode\expandafter\tilde\else\expandafter\~\fi}
\catcode248=13 \def ø{\ll}                         
\catcode249=13 \def ù{\gg}                       
\catcode250=13 \def ú{\eta}                      
\catcode251=13 \def û{\kappa}                  
\catcode252=13 \def ü{\half}     		 
\catcode253=13 \def ý{\Gamma} 		
\catcode254=13 \def þ{\Xi}   			
\catcode255=13 \def ÿ{\relax\ifmmode\expandafter{}^{\dagger}{}\else\dag\fi}

\def\S{{\cal S}}

\def\U{{\cal U}}

\title{
\color{blue}
F-theory from Fundamental Five-branes
}
\author{William D. Linch \textsc{iii}$^\text{\Pisces}$ and Warren Siegel$^\text{\Scorpio}$}
\date{}		

\begin{document}
\maketitle

\vspace*{-65mm}
\begin{flushright}
UMDEPP-015-002\\
YITP-SB-15-3
\end{flushright}
\vspace*{+40mm}

\begin{center}
{\em
${}^{\mbox{\footnotesize\Pisces}}$
Center for String and Particle Theory,
Department of Physics,\\
University of Maryland at College Park,
College Park, MD 20742-4111\\
~\\
${}^{\mbox{\footnotesize\Scorpio}}$
C. N. Yang Institute for Theoretical Physics\\
State University of New York, Stony Brook, NY 11794-3840
}
\end{center} 

\vspace{10pt}

\begin{abstract}

We describe the worldvolume for the bosonic sector of the lower-dimensional F-theory that embeds 4D, N=1 M-theory and the 3D Type II superstring.  
The worldvolume (5-brane) theory is that of a single 6D gauge 2-form $X_{MN}(\sigma^P)$
whose field strength is selfdual.
Thus unlike string theory, the spacetime indices are tied to the worldsheet ones: In the Hamiltonian formalism, the spacetime coordinates are a {\bf 10} of the GL(5) of the 5 $§$'s (neglecting $ $). 
The current algebra gives a rederivation of the F-bracket. 
The background-independent subalgebra of the Virasoro algebra gives the usual section condition, while a new type of section condition follows from Gau\ss{}'s law, tying the worldvolume to spacetime:  Solving just the old condition yields M-theory, while solving only the new one gives the manifestly T-dual version of the string, and the combination produces the usual string.
We also find a covariant form of the condition that dimensionally reduces the string coordinates.

\end{abstract}

\vspace*{.5cm}
\begin{flushleft}
~\\
{${}^{\mbox{\footnotesize\Pisces}}$ \href{mailto:wdlinch3@gmail.com}{wdlinch3@gmail.com}}\\
{$^{\text{\Scorpio}}$ \href{mailto:siegel@insti.physics.sunysb.edu}{siegel@insti.physics.sunysb.edu}}
\end{flushleft}

\setcounter{page}0
\thispagestyle{empty}

\newpage


\section{Introduction}
We continue our considerations \cite{Polacek:2014cva,Linch:2015lwa} of how string theory (S-theory) relates to its manifestly T-dual formulation (T-theory), M-theory, and the interpretation of F-theory as the manifestly STU-dual version of all these.  So far all the work on this F-theory 
\cite{
Berman:2011cg,
Berman:2011jh,
Berman:2011pe,
Godazgar:2014nqa, 
Musaev:2014lna,
Hatsuda:2013dya,
Hatsuda:2012vm,
Hatsuda:2012uk,
Park:2013gaj, 
Blair:2013gqa, 
Blair:2014zba,
Polacek:2014cva,
Linch:2015lwa
} 
has described only the massless sector.  However, T-theory was originally derived from string current algebra \cite{Siegel:1993xq, Siegel:1993th, Siegel:1993bj}.  In this paper we approach F-theory from its formulation as a fundamental 5-brane, using current algebra to derive its symmetries, and how they act on a massless background.  This fundamental 5-brane is a dynamical one, meant to be first-quantized (and maybe second) in a manner similar to the string.  (Details of the dynamics will be left for the future.)

Our treatment of fundamental branes differs from previous versions 
\cite{
Dirac:1962iy,
Collins:1976eg,
Howe:1977hp,
Sugamoto:1982hx,
Hughes:1986fa,
Bergshoeff:1987cm,
deWit:1988ig
} 
in that the worldvolume fields $X_{MN}(\sigma^P)$ describing spacetime coordinates are gauge fields.
Also, the currents are all linear in $X$, unlike treatments based on consideration of Wess-Zumino terms \cite{Hatsuda:2012uk,Hatsuda:2012vm,Hatsuda:2013dya}.  For the case of F-theory embedding the 3D string, this gauge field is the 6D gauge 2-form with selfdual field strength.  The worldvolume indices on this gauge field $X$ (actually the 5D indices remaining in a temporal gauge) thus become identified with spacetime indices.  Analysis of the current algebra of this theory naturally leads to the spacetime gauge fields of the massless sector of F-theory, their gauge transformations, the F-bracket resulting from their algebra, the F-section condition, etc.  Just as the indices of the gauge field tie the worldvolume to spacetime, so does the 2-form's Gau\ss{} law, which adds a constraint to the generalized Virasoro algebra as well as a new section condition that reduces $§$ as well as $x$.

\section{Currents and constraints}
Covariant selfdual 6D field theory has been described previously in terms of an action \cite{Siegel:1983es}.
For simplicity, we start here in a ``conformal gauge" where both the 6D metric and Lagrange multiplier for selfduality have been fixed, using the action only to define the Hamiltonian formalism, which we find convenient for our analysis.

The action is then, in Lagrangian form,
\begin{align}
S = \tfrac1{12} \int \, F_{MNP}F^{MNP} \, d^6\sigma
\end{align}
with $F_{MNP}=\tfrac12 \partial_{[P}X_{MN]}$ and $F^{(\pm)}_{MNP} = F_{MNP} \pm \tfrac1{3!} ·_{MNP}{}^{QRS} F_{QRS}$. 
The momentum conjugate to $X^{mn}$ is $P_{mn}$.
In Hamiltonian form ($\sigma^M\to \tau, \sigma^m$)
\begin{align}
S &= - \int \tfrac12 P_{mn}\partial_\tau X^{mn}  d^5\sigma d\tau  + \int H \,  d\tau \cr
H	&=  \int \left( 
	\tfrac14 P_{mn}P^{mn} 
		+ \tfrac1{12} F_{mnp}F^{mnp} 
		+ X^{\tau m} \partial^n P_{mn} 
	\right)  d^5\sigma ,
\end{align}
where the field strengths in Hamiltonian language are
\begin{align}
F_{mnp} = \tfrac 12 \partial_{[p} X_{mn]}
~~~\mathrm{and}~~~
P_{mn} = F_{\tau mn} .
\end{align}
The selfdual field strengths are the currents for the covariant derivatives \cite{Siegel:1985xj}
\begin{empheq}[box=\fcolorbox{green}{white}]{align}
\label{currents}
\dd_{mn} := F^{(+)}_{\tau mn} = P_{mn} + \tfrac12 ·_{mnpqr} \partial^r X^{pq} ,
\end{empheq}
while the antiselfdual field strengths are the symmetry currents
\begin{align}
\tilde \dd_{mn}:= F^{(-)}_{\tau mn} = P_{mn} - \tfrac12 ·_{mnpqr} \partial^r X^{pq} .
\end{align}

As usual (cf.¼electromagnetism) the time components of the gauge field $X$ become Lagrange multipliers. 
After using them to identify the constraint (Gau\ss{}'s law), we eliminate them by choosing a temporal gauge.
The Virasoro algebra is then defined by the energy-momentum tensor for the selfdual field strength 
\begin{align}
\mathcal T_{MN} = \tfrac14 F_M{}^{PQ} F_{NPQ}
~~~\Rightarrow~~~
\mathcal T^{(+)}_{MN} = \tfrac18 F^{(+)}_M{}^{PQ} F^{(+)}_{NPQ}.
\end{align}
(The unusual normalization is consequence of our definition of $F_{MNP}^{(+)}$.)
This is symmetric and traceless. Its Hamiltonian components $(\mathcal T^{(+)mn}, \mathcal T^{(+)\tau m}, \mathcal T^{(+)\tau \tau})$ are
\begin{align}
\label{E:constraints}
\mathcal T_{mn} = \eta_{mn} \mathcal T - \tfrac12 \eta^{pq} \dd_{mp}\dd_{nq}
	~,~~
\mathcal S^r = \tfrac1{16} ·^{rmnpq} \dd_{mn}\dd_{pq}
	~,~~
\mathcal T = \tfrac18 \eta^{mp} \eta^{nq} \dd_{mn}\dd_{pq}	.
\end{align}
The Gau\ss{} constraint is
\begin{align}
\mathcal U_m := \partial^n P_{mn} .
\end{align}

For purposes of analyzing kinematics, we need consider only those constraints that are GL(5) covariant, that is, need not involve the SO(3,2) metric $ú_{mn}$.  (This requires treating contravariant indices on $X^{mn}$ as opposite to those on $\sigma_m$.)
This is the subset that's background independent, since the background is introduced as a GL(5)/SO(3,2)GL(1) element to break GL(5) to SO(3,2).
Various section-like conditions can then be introduced by replacing some string coordinates in these constraints with their zero-modes \cite{Kugo:1992md}:
\begin{subequations}
\label{subsections}
\setlength{\fboxsep}{10pt}
\begin{empheq}[box=\fcolorbox{green}{white}]{align}
& \hbox{Virasoro} & \mathcal S^m =¼& \tfrac1{16} ·^{mnpqr} \dd_{np}\dd_{qr} && \\
& \hbox{dimensional reduction} & \on\circ\S{}^m :=¼& \f14 ·^{mnpqr}p_{np}P_{qr} & \mathcal U_m =¼& \partial^n P_{mn} \\
\label{subsection}
& \hbox{section condition} & \under\circ\S{}^m :=¼& \f18 ·^{mnpqr}p_{pq}p_{qr} & \under\circ\U{}_m :=¼& »^n p_{mn}
\end{empheq}
\end{subequations}
where the section conditions are to be interpreted as being applied as both
\begin{equation}
AB = 0âÜâABf = (Af)(Bg) = 0
\end{equation}
for $f,g$ that are functions of $§$ or $X(§)$ (for hitting with $»^m$) or functions of $x$ or $X(§)$ (for hitting with $p_{mn}$).

We thus have 3 types of conditions:
\begin{enumerate}
\item We generalized (the background independent part of) the string Virasoro algebra with the generators $\S^m$ of coordinate transformations for the 5 $§$'s.  
\item We treat Gau\ss{}'s law $\U_m$, which arises because the 6D $X$ is a gauge field, as a dimensional reduction condition since it's linear in the string variables. We also have a new covariant dimensional reduction condition $\on\circ\S{}^m$.  (It simplified using $\under\circ\U{}_m$.  
Since $»^{[m}»^{n]}=0$, both dimensional reduction conditions can be written with $P$ replaced with either $\dd$ or $\tilde\dd$:  The latter allows them to commute with Virasoro.)  It replaces $P_L-P_R$ used in the manifestly T-dual version of the string that has doubled coordinates.  (This reduces to $ú^{mn}p_m P_n$ in that formalism.  Dimensional reduction for doubled coordinates was invented in \cite{Siegel:1994xr}.  T-theory with doubled selfdual scalars was considered in \cite{Hohm:2013jaa}.  T-theory with both selfdual and anti-selfdual scalars was attempted in \cite{Tseytlin:1990nb, Tseytlin:1990va}.)  
\item The section conditions include $\under\circ\S$, originally found by closure of gauge transformations (see below) in F-gravity \cite{Berman:2011cg}, and a new one $\under\circ\U$ that mutually restricts $x$ and $§$.  (The former condition reduces to the original section condition $üú^{mn}p_m p_n$ in T-theory \cite{Siegel:1993xq, Siegel:1993th, Siegel:1993bj}.)
\end{enumerate}

\section{Algebras and gauge symmetries}
Using the Poisson bracket 
\begin{align}
\left[ P_{mn}(1) , X^{pq}(2) \right] = - i \delta^p_{[m} \delta^q_{n]} \delta(1 - 2) ,
\end{align}
we find the algebra
\begin{align}
\label{E:CurrentAlgebra}
\left[ \dd_{mn}(1) , \dd_{pq}(2) \right] 
	&= 2i ·_{mnpqr} \partial^r\delta(1-2) \cr
\left[ \dd_{mn}(1) , \tilde \dd_{pq}(2) \right] 
	&= 0\cr
\left[ \tilde  \dd_{mn}(1) , \tilde \dd_{pq}(2) \right] 
	&= -2i ·_{mnpqr} \partial^r\delta(1-2) .
\end{align}
Selfduality ($\tilde \dd = 0$) is thus a second-class constraint (as for string scalars \cite{Siegel:1985xj}), but we saw above the covariant division into first-class as $\f14 ·^{mnpqr}p_{np}P_{qr}=\f14 ·^{mnpqr}p_{np}\tilde\dd_{qr}$ (using a section condition).
Unlike \cite{Siegel:1983es}, the new constraint is linear in the string coordinates.

From the $\dd\dd$ commutation relations we find
\begin{align}
\label{E:SU>}
\left[\mathcal S^p(1) , \dd_{mn}(2) \right] 
	&= \tfrac i2 \partial^q \delta(1-2)  \delta^p_{[q} \dd_{mn]}(1) 
~~~\mathrm{and}~~~
\left[ \mathcal U_{p}(1) , \dd_{mn}(2) \right] 
	= 0	.
\end{align}
Straighforward calculation gives
\begin{align}
\label{E:braneAlgebra}
\left[ \mathcal S^m(1) , \mathcal S^n(2) \right] 
	=¼& i \partial^{(m} \delta(1-2) \mathcal S^{n)}\tfrac12((1)+(2))
		- \tfrac i2 \delta(1-2) \left[ \partial^{[m} \mathcal S^{n]} - ·^{mnpqr} \dd_{pq}\mathcal U_r\right]
	\cr
[\S, \U ] = [\U, \U ] =¼& 0
\end{align}
where we are defining $\mathcal O\tfrac12((1)+(2)):= \tfrac12\left[\mathcal O (1) +\mathcal O (2)\right]$.

On functions $f(X) = f(0) + \tfrac 12 X^{mn}\partial_{mn} f(0) + \mathcal O(X^2)$, the Poisson brackets with the currents give spacetime derivatives
\begin{align} 
[ \dd_{mn}(1) , f(X(2)) ] 
	= 
	-i \partial_{mn} f \delta(1-2)
	=
[ \tilde \dd_{mn}(1) , f(X(2)) ].
\end{align}
The worldvolume version of this is given by
\begin{align}
 [(\mathcal S^m- \tilde {\mathcal S}^m)(1) , f(2) ] = -i \partial^m f \delta(1-2)
\end{align}
(with $\tilde {\mathcal S}$ formally the same as ${\mathcal S}$ but with $\dd$ replaced by $\tilde \dd$) so that, up to second-class constraints, ${\mathcal S}$ generates translations in $\sigma$.
To see this, note that 
\begin{align}
\label{E:partials}
\partial^r f &= 
	\tfrac18 ·^{mnpqr} (\partial_{mn} f) (\dd_{pq} -\tilde \dd_{pq})
	-(\partial^m X^{nr})(\partial_{mn} f)
=  \tfrac18 ·^{mnpqr} (\partial_{mn} f) (\dd_{pq} -\tilde \dd_{pq})
\end{align}
modulo 
the new section condition (\ref{subsection})
\begin{align}
\label{E:section2}
(\partial^n f) (\partial_{mn}g) = 0
~~~\forall ~ f(X), g(X).
\end{align} 
(In T-theory the analogue of this formula for $\partial f$ was derived by dimensionally reducing (solving the second-class constraint) and then oxidizing \cite{Siegel:1993th}.) 

Worldvolume ($§$) reparameterizations are generated by 
\begin{align}
\delta_\xi = i \int \xi_m \, \mathcal S^m. 
\end{align}
Then (\ref{E:SU>}) implies
\begin{align}
\delta_\xi \dd_{mn} 
	&= \tfrac12 \partial^p (\xi_{[p} \dd_{mn]}) 
	= (\partial^p \xi_p) \dd_{mn} +(\partial^p \xi_{[m}) \dd_{n]p}
		+ \xi_{p} \partial^p \dd_{mn} + \xi_{[m} \partial^p\dd_{n]p} ,
\end{align}
corresponding to a density term (integrable on scalars), two contravariant index transformation terms, a coordinate transformation, and two terms for integrability on non-scalars. This demonstrates that $\dd$ isn't exactly a tensor, but its integral is invariant. 

Using (\ref{E:partials}), we can compute the commutator of two vector fields $V_i^{mn}$ for ${}_i={}_1,{}_2$:
\begin{align}
\label{E:master}
\left[ V_1^{mn} \dd_{mn} , V_2^{pq} \dd_{pq}\right] &=
	2i\epsilon_{mnpqr} \partial^r \delta(1-2) V_1^{mn}V_2^{pq}\tfrac12((1)+(2)) 
\cr
	&-i \delta(1-2) \left[ V_{[1}^{mn}\partial_{mn} V_{2]}^{pq} -\tfrac18 V_{[1}^{[mn}\partial_{mn}V_{2]}^{pq]} \right] \dd_{pq},
\end{align}
modulo the second-class constraints (and sectioning). We use this to check the algebra of (spacetime) gauge transformations: Defining
\begin{align}
\Lambda_i := \tfrac i2 \int \lambda_i^{mn} \dd_{mn} 
\end{align}
for ${}_i={}_1,{}_2$, we find for their commutator
\begin{align}
\left[ \Lambda_1, \Lambda_2\right] 
	&= \tfrac i2 \int \left[ \tfrac12 \lambda_{[1}^{mn}\partial_{mn} \lambda_{2]}^{pq} \dd_{pq}
		-\tfrac 1{16} \lambda_{[1}^{[mn}\partial_{mn} \lambda_{2]}^{pq]} \dd_{pq}\right]
.
\end{align}
This shows that the algebra of gauge transformations (cf. ref. \cite{Berman:2011cg}) closes
\begin{align}
\label{E:vf}
\left[ \Lambda_1 , \Lambda_2 \right] = \Lambda_{12}
~~~\mathrm{with}~~~
\lambda_{12}^{mn} &= \tfrac12 \lambda_{[1}^{pq}\partial_{pq} \lambda_{2]}^{mn} 
	- \tfrac1{16} \lambda_{[1}^{[mn}\partial_{pq} \lambda_{2]}^{pq]}
\cr
	&= \tfrac14 \lambda_{[1}^{pq}\partial_{pq} \lambda_{2]}^{mn}
	-\tfrac12 \lambda_{[1}^{p[m}\partial_{pq} \lambda_{2]}^{n]q}
	-\tfrac14 \lambda_{[1}^{mn}\partial_{pq} \lambda_{2]}^{pq} .
\end{align}
Note that (\ref{E:partials}) implies that the gauge parameter itself has the gauge ambiguity 
\begin{align}
\delta \lambda^{mn} = \tfrac12 \epsilon^{mnpqr}\partial_{pq}\lambda_r.
\end{align}

\section{Backgrounds}

Background fields are introduced as usual through the covariant derivatives
\begin{equation}
\dd_{ab} = \tfrac12 E_{ab}{}^{mn}\dd_{mn} .
\end{equation}
Then (\ref{E:master}) applied to $\delta_\lambda \dd_{ab} = [\Lambda, \dd_{ab}]$ gives
\begin{align}
\label{E:deltaE}
\delta_\lambda E_{ab}{}^{mn} 
	&= \tfrac12 \lambda^{pq}\partial_{pq} E_{ab}{}^{mn}
	 -\tfrac12 E_{ab}{}^{pq}\partial_{pq} \lambda^{mn}
	+\tfrac18 E_{ab}{}^{[mn}\partial_{pq} \lambda^{pq]} 
\cr &= \tfrac12 \lambda^{pq}\partial_{pq} E_{ab}{}^{mn}
	+\tfrac12 E_{ab}{}^{mn}\partial_{pq} \lambda^{pq}
	+ E_{ab}{}^{p[m}\partial_{pq} \lambda^{n]q} ,
\end{align}
in agreement with \cite{Berman:2011cg}.
(There is also a nonderivative Sp(4) gauge transformation on the flat indices $a,b$.)

Factorization of the vielbein follows from requiring that the result $\S^a$ of replacing $\dd_{mn}$ with $\dd_{ab}$ in $\S$ does not generate an independent symmetry.   This is essentially the statement that $·$ is a tensor under transformation by the vielbein, and thus the vielbein is an element of GL(5):  Introducing linear dependence through a new vielbein $E_m{}^a$,
\begin{align}
\mathcal S^a \sim \mathcal S^m E_m{}^a
~~\Rightarrow~~
\f14 ·^{abcde} E_{bc}{}^{mn} E_{de}{}^{pq} \sim ·^{mnpqr}E_r{}^a
~~\Rightarrow~~
E_{ab}{}^{mn} = E_{[a}{}^m E_{b]}{}^n
\end{align}
where $E_a{}^m$ is the inverse of $E_m{}^a$, and we have chosen the proportionality factor to be det($E_a{}^m$) in the final step for convenience.  Thus $E_a{}^m$ and $E_{ab}{}^{mn}$ are representations of GL(5) in the {\bf 5} and {\bf 10} representations, each of which has been expressed in terms of the other above.  

Alternatively, using (\ref{E:master}) again, one computes that
\begin{align}
\label{E:>>flat}
[\dd_{ab}(1), \dd_{cd}(2)] 
&= \tfrac i2 \epsilon_{mnpqr} \partial^r \delta(1-2)
			E_{ab}{}^{mn}E_{cd}{}^{pq}\tfrac12((1)+(2))
\\		
& -\tfrac i8 \delta(1-2) \left[
			E_{ab}{}^{mn}\partial_{mn} E_{cd}{}^{pq} 
			-\tfrac18 E_{ab}{}^{[mn}\partial_{mn} E_{cd}{}^{pq]} 
			-{}_{ab} \leftrightarrow {}_{cd}
			\right]  E_{pq}{}^{ef}\dd_{ef} .
\nonumber
\end{align}
Then the first term has to be proportional to $\epsilon_{abcde} E_m{}^e \partial^m \delta(1-2)$.
Since $E_a{}^m$ is thus an unconstrained matrix, it's more convenient to use as the fundamental field. We therefore use (\ref{E:deltaE}) to find its gauge transformation:
\begin{align}
\delta E_a{}^m &= 
	\tfrac12 \lambda^{pq} \partial_{pq} E_a{}^{m} 
	+\tfrac14 E_a{}^m \partial_{pq}\lambda^{pq}
	+E_a{}^{p} \partial_{pq}\lambda^{qm} .
\end{align}
To rewrite (\ref{E:>>flat}) in terms of the fundamental field, it is useful to flatten the indices on the derivatives 
$\partial_{ab} := E_a{}^m E_b{}^n \partial_{mn}$. The anholonomy coefficients 
\begin{align}
[\partial_{ab}, \partial_{cd}] 
	= \tfrac12 \left[ c_{ab\,cd}{}^{ef}  - c_{cd\,ab}{}^{ef} \right] \partial_{ef}
\end{align}
reduce to $c_{ab\, cd}{}^{ef} = (\partial_{ab} E_{[c}{}^m)E_m{}^{[e} \delta_{d]}^{f]} =: c_{ab \, [c}{}^{[e}\delta_{d]}^{f]}$.
In terms of these, (\ref{E:>>flat}) becomes
\begin{align}
[\dd_{ab}(1), \dd_{cd}(2)] 
	&= \tfrac i2 \delta(1-2) \left[		
		c_{e[a\, b]}{}^e \dd_{cd}
		+c_{ab\, [c}{}^e \dd_{d]e} 
		+ c_{e[c|\,[a}{}^e \dd_{b]|d]}
		+ c_{[c|[a\, b]}{}^e \dd_{|d]e}		
		- ({}_{[ab]} \leftrightarrow {}_{[cd]}) \right]  \nonumber\\
	&+2i \mathrm{det}(E_a{}^m) \epsilon_{abcde} \partial^e \delta(1-2),
\end{align}
where $\partial^a := E_m{}^a \partial^m$ and we have rewritten $E_a{}^mE_b{}^nE_c{}^pE_d{}^qE_e{}^r\epsilon_{mnpqr} = \mathrm{det}(E_a{}^m) \epsilon_{abcde}$.
%

\section{Sections: F \texorpdfstring{$\to$}{\textrightarrow} M,T,S}
Sectioning is straightforward.  Solving the old section condition $\under\circ\S{}^m$ as before, but also the new dimensional reduction condition $\on\circ\S{}^m$, gives (for $m=-1,0,1,2,3$)

\begin{align}
·^{mnpqr}p_{np}p_{qr} = 0â&Üâp_{ij} = 0¼;âp = p_{-1i}â(i=0,1,2,3) \cr 
·^{mnpqr}p_{np}P_{qr} = 0â&ÜâP_{ij} = 0¼;âP = P_{-1i} \cr 
\dd_{-1i} = P_{-1i}¼&,â\dd_{ij} = -·_{ijkl}»^k X^{-1l} \cr 
\S^i = (»^{[i} X^{-1j]})P_{-1j}¼&,â\S^{-1} = ü·_{ijkl}(»^i X^{-1j})(»^k X^{-1l}) \cr 
\U_i = -»^{-1}P_{-1i}¼&,â\U_{-1} = »^i P_{-1i}
\end{align}
describing M-theory, still on a 5-brane, but in 4 spacetime dimensions.

On the other hand, solving the new conditions $\U$ and $\under\circ\U$ gives
\begin{align}
»^n p_{mn} = 0â&Üâ»^i = p_{3i} = 0¼;â» = »^3¼,âp = p_{ij}â(i=-1,0,1,2)\nonumber\\
»^n P_{mn} = 0â&ÜâP_{3i} = 0¼;âP = p_{ij}\nonumber\\
\dd_{3i} = 0¼&,â\dd_{ij} = P_{ij} +ü·_{ijkl}»^3 X^{kl}\nonumber\\
\S^i = 0¼&,â\S^3 = \f18 ·^{ijkl}\dd_{ij}\dd_{kl}\nonumber\\
\on\circ\S{}^i = 0¼&,â\on\circ\S{}^3 = \f14 ·^{ijkl}p_{ij}P_{kl}
\end{align}
describing T-theory on a 1-brane (string), in 6 (i.e., doubled) dimensions.

Solving both sets of conditions gives S-theory:
\begin{align}
» = »^3¼,â& p = p_{-1i}â(i=0,1,2)\nonumber\\
& P = P_{-1i}\nonumber\\
\dd_{3i} = 0¼;â& \dd_{-1i} = P_{-1i}¼,â\dd_{ij} = -·_{ijk}»^3 X^{-1k}\nonumber\\
\S =¼&\S^3 = (»^3 X^{-1i})P_{-1i}
\end{align}

\section{Conclusions}
Starting from the worldvolume theory of a selfdual gauge form in D=6, we have derived the conditions (\ref{subsections}) that generalize string theory to F-theory.  All of them are new except for the section condition of F-gravity, which we have now found from first principles, along with the field representation of F-gravity.  (Reduction to T-theory also gives a new covariant form for its dimensional reduction condition.)  Their algebra follows from that of the (selfdual) currents (\ref{currents}), which generate the gauge transformations of F-gravity.  (The constraints $\S^m$ generate $§$ reparametrizations.)

Many future avenues of investigation are suggested: 
\begin{enumerate}
	\item The covariant 6D conformal field theory might be useful, e.g., for analyzing $Œ'$ corrections.  This would require an analysis of the algebra of the ``Ê$\mathcal T$Ê" constraints in (\ref{E:constraints}) responsible for the dynamics.  
A related problem is the 6D worldvolume vielbein: If part is identified with the spacetime vielbein, then the remainder might be the 6 gauge fields/Lagrange multipliers for worldvolume coordinate transformations.
	\item Including the currents for SO(3,2) (or SO(3,3) for the 6D formulation) would allow the definition of truly covariant derivatives on the worldvolume and their torsion and curvature \cite{Polacek:2013nla}.
	\item We have looked at only the bosonic string. Generalization to the superstring should be straightforward using \cite{Linch:2015lwa}.
Then reduction to T-theory can be compared to the formulation with
Ramond-Ramond currents \cite{Hatsuda:2014aza}. The tying of worldvolume and spacetime symmetries in the bosonic case suggests that the Green-Schwarz and Ramond-Neveu-Schwarz formulations might be more directly related.
	\item Of course, these results should be generalized to higher dimensions. However, simple use of selfdual forms would give different cosets than those found in the bosonic sectors of supergravities. Supersymmetry, especially for the D = 4, 6, and 10 superstrings, should place new restrictions. 
For the D = 4 case, the bosonic coordinates are a spinor {\bf 16} representation of SO(5,5).ÊÊThen the section condition uses a 10D $\gamma$-matrix \cite{Coimbra:2011ky, Berman:2012vc}.  The bosonic covariant derivative then resembles a fermionic supersymmetry-covariant derivative:
\begin{equation}
\dd_µ = P_\mu + (\gamma_m)_{\mu\nu} \partial^m X^\nu¼Ü¼
[\dd_\mu,\dd_\nu] = 2i (\gamma_m)_{\mu \nu} \partial^m \delta¼.
\end{equation}
Also, 
\begin{equation}
\S^m = \tfrac14 (\gamma^m)^{\mu \nu} \dd_\mu \dd_\nu¼.
\end{equation}
\end{enumerate}

\section*{Acknowledgements}
W{\sc dl}3 thanks Brenno Carlini Vallilo for discussions. W{\sc dl}3 is partially supported by the U{\sc mcp} Center for String \& Particle Theory and National Science Foundation grants PHY-0652983, and PHY-0354401. 
W{\sc s} is supported in part by National Science Foundation grant PHY-1316617. 


\small
\baselineskip=15pt
\bibliography{/Users/wdlinch3/Dropbox/Rashoumon/BibTex}
\bibliographystyle{unsrt}

\end{document}